\documentclass[sigconf]{acmart}

\usepackage[T1]{fontenc}
\usepackage{amsmath,amsfonts}
\usepackage{algorithm}
\usepackage[noend]{algorithmic}
\usepackage{graphicx}
\usepackage{textcomp}
\usepackage{xcolor}
\usepackage{balance}
\usepackage{todonotes}
\usepackage{url}
\usepackage{eqparbox}

\DeclareMathOperator*{\argmin}{argmin}
\newcommand*{\argminl}{\argmin\limits}

\AtBeginDocument{%
	  }

\setcopyright{none}
\copyrightyear{2025}
\acmYear{2025}
\acmDOI{XXXXXXX.XXXXXXX}
\acmConference[14th International ParBio Workshop  '25]{14th International Workshop on Parallel and AI-based Bioinformatics and Biomedicine}{October 12, 2025}{Philadelphia, PA}
\acmISBN{978-1-4503-XXXX-X/2018/06}

\begin{document}

\title{Optimizing the Variant Calling Pipeline Execution on Human Genomes Using GPU-Enabled Machines}

\author{Ajay Kumar}
\affiliation{%
	\institution{The University of Missouri}
     \city{Columbia}
 	\country{USA}}
\email{ajay.kumar@missouri.edu}

\author{Praveen Rao}
\affiliation{%
	\institution{The University of Missouri}
	\city{Columbia}
	\country{USA}}
\email{praveen.rao@missouri.edu}

\author{Peter Sanders}
\affiliation{%
	\institution{Karlsruhe Institute of Technology}
	\city{Karlsruhe}
	\country{Germany}}
\email{sanders@kit.edu}


%
%


\begin{abstract}
Variant calling is the first step in analyzing a human genome and aims to detect variants in an individual's genome compared to a reference genome. Due to the computationally-intensive nature of variant calling, genomic data are increasingly processed in cloud environments as large amounts of compute and storage resources can be acquired with the pay-as-you-go pricing model. In this paper, we address the problem of efficiently executing a variant calling pipeline for a workload of human genomes on graphics processing unit (GPU)-enabled machines. We propose a novel machine learning (ML)-based approach for optimizing the workload execution to minimize the total execution time. Our approach encompasses two key techniques: The first technique employs ML to predict the execution times of different stages in a variant calling pipeline based on the characteristics of a genome sequence. Using the predicted times, the second technique generates optimal execution plans for the machines by drawing inspiration from the flexible job shop scheduling problem. The plans are executed via careful synchronization across different machines. We evaluated our approach on a workload of publicly available genome sequences using a testbed with different types of GPU hardware. We observed that our approach was effective in predicting the execution times of variant calling pipeline stages using ML on features such as sequence size, read quality, percentage of duplicate reads, and average read length. In addition, our approach achieved $2\times$ speedup (on an average) over a greedy approach that also used ML for predicting the execution times on the tested workload of sequences. Finally, our approach achieved $1.6\times$ speedup (on an average) over a dynamic approach that executed the workload based on availability of resources without using any ML-based time predictions.
\end{abstract}



\maketitle

\section{Introduction}
Today, whole genome sequencing (WGS) is routinely used in large-scale genomic studies and clinical practice~\cite{WGSClinicalPractice2024} due to its economic feasibility\footnote{www.genome.gov/about-genomics/fact-sheets/sequencing-human-genome-cost}. In recent years, new genomic initiatives have emerged for the diagnosis and treatment of life-threatening diseases such as cancer and COVID-19\footnote{https://www.covidhge.com}. For example, in the UK\footnote{https://www.ukbiobank.ac.uk} and the USA\footnote{https://allofus.nih.gov}, researchers have already sequenced 500K and 250K whole genome sequences of individuals, respectively. As projected by Stephens et al.~\cite{GenomicsBigData2015}, the volume of human genome data is growing rapidly globally. Hence, the processing and analysis of massive genomic data continues to pose new challenges.   



\textit{Variant calling} is the first step performed on an individual's genome to identify variants such as single nucleotide polymorphisms (SNPs), short insertions/deletions (indels), and copy number variations compared to a reference genome. These variants play a key role in assessing an individual's risk for diseases such as cancer and the development of treatment options. A variant calling pipeline~\cite{Koboldt2020} is a software executed to identifying such variants and consists of \textit{several stages}. The stages include reading a raw genome sequence, performing alignment of the deoxyribonucleic acid (DNA) fragments (a.k.a reads) in it with a reference genome (e.g., GRCh38~\cite{GRCh38}, the Human Pangenome Reference~\cite{liao2023draft}), additional pre-processing steps for correcting sequencing errors, and invoking a variant calling method~\cite{GATK4,poplin2018universal}. The process is \textit{computationally intensive} as a genome sequence is large in size (i.e., 10 to 100+ gigabytes (GB)) due to billions of base pairs in the human DNA~\cite{committee1988mapping} and millions of reads (including overlapping reads at every position in the sequence) for correcting sequencing errors. Clinical-grade sequences have high number of overlapping reads (or coverage) at every position (e.g., 30$\times$ coverage). Fortunately, cloud computing enables users to provision (on-demand) compute/storage resources with the pay-as-you-go pricing model and is a feasible way of analyzing large genome workloads~\cite{CloudGenomics2018,SRACloud2022}.

There is continued interest in accelerating variant calling pipelines using distributed computing and hardware accelerators~\cite{SIGMODADAM15,Nothaft2017,GATK4,RaoAVAH2021,RaoAVAH*2023}. On the commercial front, companies are developing new software and services for human genome analysis. Microsoft Genomics, AWS HealthOmics, Google's Cloud Life Sciences, and Terra support cloud-based processing of genomic workflows. In fact, NVIDIA Parabricks~\cite{OConnell2022} is a free software developed for accelerating best practice pipelines (e.g., GATK~\cite{GATK4}, DeepVariant~\cite{poplin2018universal}) using GPUs. Parabricks was up to 65$\times$ faster on GPUs (compared to CPUs) for different variant callers. As the software ecosystem and the size of genome workloads continue to grow, it is timely and critical to further improve the performance of variant calling on large genome workloads while reducing the processing cost.


Motivated by the aforementioned reasons, we propose a new ML-based approach for \textit{optimizing the execution of a variant calling pipeline} for a genome workload on GPU-enabled virtual machines (VMs). The key contributions of our work are as follows:
\begin{itemize}
\item
Our first contribution is to demonstrate that ML can effectively predict the execution time of different variant calling pipeline stages for a genome sequence in the workload. This is necessary to formulate the generation of execution plans for different VMs as an optimization problem. In addition to the size of a genome sequence, the ML models employ additional characteristics of a sequence (e.g., number of bases, average read length, unique reads, read quality) to define the model features. We trained ML models such as Random Forest (RF), XGBoost\footnote{https://xgboost.ai}, Lasso Regression (LR), Support Vector Machine (SVM), and neural networks (NN). The training data was generated by executing a variant calling pipeline software on publicly available WGS data using different GPU-enabled VMs. 
\item
Our second contribution is the generation of optimal execution plans (using the predicted times) by drawing inspiration from the flexible job shop scheduling problem (FJSP)~\cite{FJSP2024}, a combinatorial optimization problem that has been extensively studied in operations research. The plans are executed via careful synchronization across different VMs (e.g., using file locking on shared storage).
\item 
We evaluated our approach on a workload of publicly available WGS data using Parabricks and five VMs with different GPU capabilities (i.e., number/type of GPUs). All ML models achieved better R$^2$ score for predicting the execution times of variant calling pipeline stages when genome sequence characteristics were used as features instead of just the sequence size. The best R$^2$ score was achieved by RF (e.g., 0.92 for 1-stage pipeline). 
\item
Furthermore, our approach (based on the FJSP) achieved $2\times$ speedup (on an average) over a greedy approach on the tested workload. The greedy approach used the ML-based time predictions. We also evaluated our approach against a dynamic approach that did not use any ML-based time predictions. The assignment of sequences to VMs for execution was dynamic based on the availability of VMs. Our approach achieved $1.6\times$ speedup (on an average) over the dynamic approach on the tested workload.
\end{itemize}


The rest of the paper is organized as follows: Section~\ref{sec-background} provides background and motivation of our work. Section~\ref{sec-approach} introduces our approach; Section~\ref{sec-evaluation} reports the performance evaluation results; and finally, we conclude in Section~\ref{sec-conclusion}.

\section{Background and Motivation}
\label{sec-background}

\subsection{Variant Calling Pipelines}


We discuss closely related work on accelerating single sample variant calling pipelines. With the availability of big data technologies such as Apache Hadoop~\cite{Hadoop} and Spark~\cite{Zaharia2010}, efforts were made to accelerate the DNA variant calling pipelines using these technologies. Some researchers accelerated the alignment stage of variant calling using Apache Hadoop~\cite{Schatz2009,Nguyen2011,SEAL2011,BigBWA2015} and Apache Spark~\cite{SparkBWA2016,Cong2015CSBWAMEMA,PipeMEM}. Another effort used Hadoop-based parallel I/O~\cite{HadoopBAM2012} for faster access to sequencing data. Hardware accelerators such as field-programmable gate arrays (FPGAs) have also been used to speed up alignment~\cite{ahmed2015heterogeneous,ChenFPGA2016} and other computationally intensive stages~\cite{LoFPGABQSR2020,XuFPGALoFreq2023}.


Next, we discuss related work that attempted to accelerate the entire variant calling pipeline. An early effort parallelized variant calling by splitting the sequences by population and chromosome on Amazon Web Services (AWS) for low-cost variant calling on 2,500+ genome sequences~\cite{Shringarpure2015}. Other variant calling pipeline software~\cite{Halvade2015,GATK4,Nothaft2017,SIGMODADAM15,huang2020deepvariant,RaoAVAH2021,RaoAVAH*2023} employed big data tools (e.g., Apache Hadoop\footnote{https://hadoop.apache.org}/Spark\footnote{https://spark.apache.org}) for parallelization; a few of the efforts focused on improving cluster utilization via asynchronous computations~\cite{RaoAVAH2021,RaoAVAH*2023}. NVIDIA developed Parabricks to accelerate GATK pipelines using GPUs~\cite{NVIDIAGenomics,OConnell2022}. Google developed DeepVariant~\cite{DeepVariant,DeepVariant3} that used deep learning for variant calling and operated directly on aligned reads. Another effort~\cite{ParallelDeepVariant} parallelized and accelerated DeepVariant~\cite{DeepVariant2} to leverage GPUs. More recently, Illumina developed the DRAGEN Platform to accelerate the variant calling pipeline using FPGAs~\cite{DRAGEN_Scheffler2023}. Sentieon\footnote{https://www.sentieon.com} developed optimized software-based algorithms for variant calling using CPUs for cloud environments. They also developed an ML-based variant caller called DNAscope~\cite{DNAscope}.

Workflow management systems such as Swift/T\footnote{https://github.com/ncsa/Swift-T-Variant-Calling}, Nextflow\footnote{www.nextflow.io}, Cromwell\footnote{https://github.com/broadinstitute/cromwell} with Common Workflow Language (CWL)/Workflow Description Language (WDL) were explored for variant calling pipelines on different computing infrastructures such as local, high performance computing (HPC) clusters, and in cloud environments~\cite{AhmedDesign2021,SPISAKOVA2023}. Mulone et al.~\cite{Mulone2023} developed a cloud-HPC hybrid workflow to lower the cost of variant calling and improve performance.


Workflow scheduling has been studied in cloud environments with the goal of minimizing makespan or makespan and cost~\cite{durillo2014multi,rodriguez2017budget,taghinezhad2022qos,xia2023cost,Silva2024}. Closely related to our work is that of Silva et al.~\cite{Silva2024}, which processes multiple linear workflows and aims to minimize makespan. However, they assume that the task times are unknown and hence, employ work stealing. On the contrary, we aim to employ ML for predicting the task execution times to generate optimal schedules.

None of the prior efforts has attempted \textit{to formulate the execution of a variant calling pipeline on a large genome workload} as an optimization problem for effective use of cluster resources such as GPUs. Our work addresses this critical gap in cloud-based variant calling to ultimately enable users to lower the cost.

\subsection{Background on the FJSP}

We describe the classical FJSP~\cite{brucker1990job,FJSP2024}, which is an NP-hard optimization problem. For a set of jobs, the goal of FJSP is to find an optimal schedule to execute them on a given set of machines. Each job can have one or more operations that must be executed in a sequence. An operation (of a job) has the choice of executing on a subset of machines with different execution times. Each machine can execute only one operation at a time, and an operation cannot be preempted. All machines and jobs are available at time 0. The goal of the FJSP is to find the optimal schedule of jobs (and their operations) on the machines to minimize an objective function. The most common criterion is to minimize the \textit{makespan}, which is the total time taken from the start of the execution of the first job till the last job completes. The FJSP can be solved using a mixed integer linear program, disjunctive graph model, or constraint programming (CP) model~\cite{FJSP2024}. 




\subsection{Motivation}

Given a workload of human genomes and GPU-enabled VMs, our goal is \textit{to minimize the makespan} so that the cost of genome processing can reduced especially in cloud environments. Randomly distributing sequences across the VMs is not ideal as the execution time of variant calling depends on a sequence's characteristics (e.g., size) and the underlying VMs' capabilities. This may lead to load imbalance and lower utilization of VM resources (e.g., GPUs) resulting in a suboptimal makespan. Hence, there are few challenges that must be addressed to develop an effective solution: The first challenge is to formulate the generation of optimal execution plans for the given genome workload and GPU-enabled VMs as \textit{an optimization problem}. This would require the execution time estimates of different variant calling pipeline stages to solve the optimization. Hence, the second challenge is \textit{to predict the execution time} of a variant calling pipeline stage on different VMs with good accuracy. Once the execution plans are generated, the final challenge is \textit{to execute them correctly} to ensure that different stages of variant calling for a genome are executed in the right order and possibly across different VMs. To the best of our knowledge, none of the prior efforts has addressed these challenges for large genome workloads using GPU-enabled VMs.

\section{Our Approach}
\label{sec-approach}

We begin by presenting our computing model for variant calling. We then introduce our new approach to innovatively tackle the aforementioned challenges. 

\subsection{Computing Model}

\begin{figure}[tbh]
\begin{center}
\begin{tabular}{c}
\includegraphics[width=2.6in, angle=0]{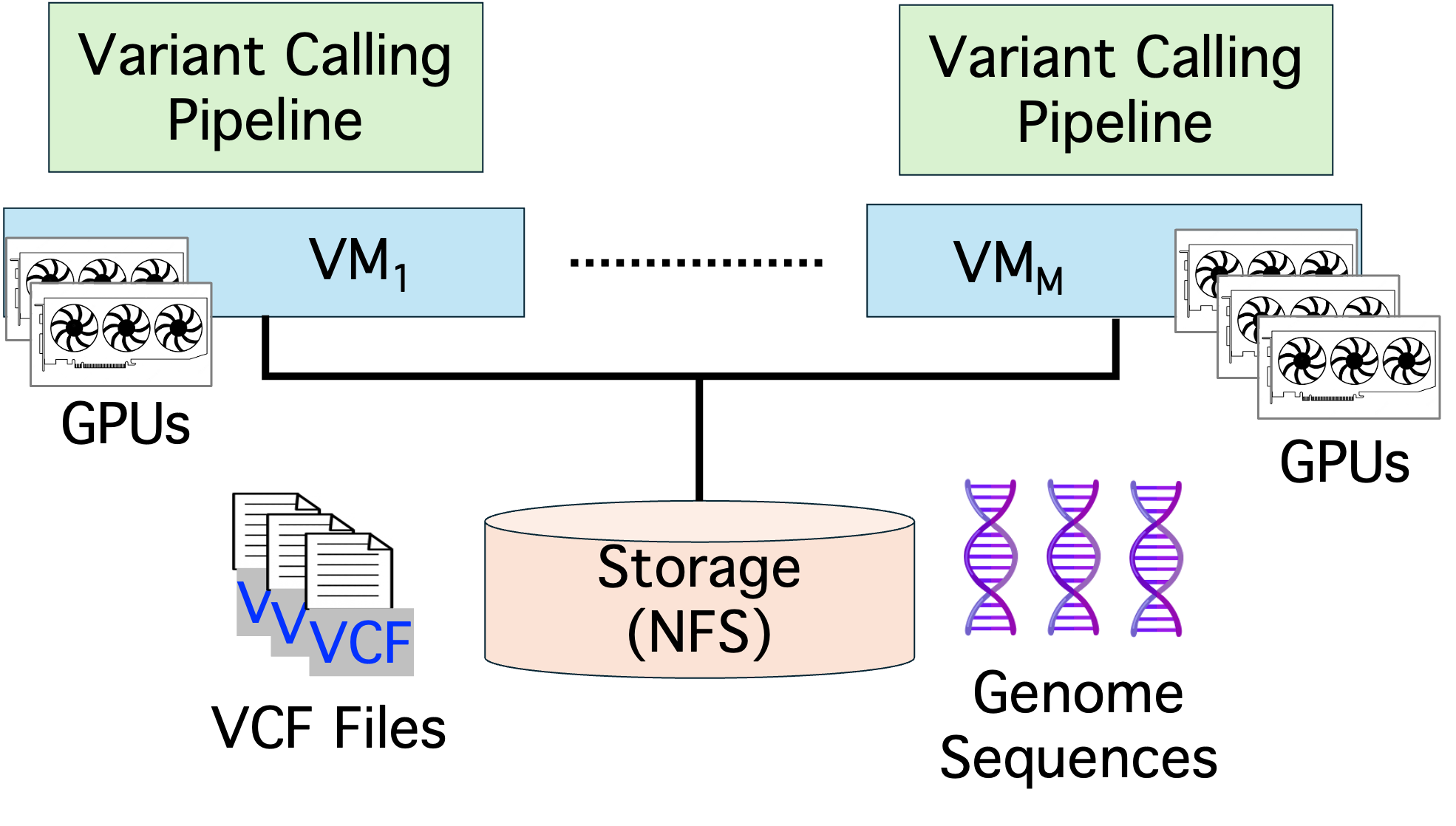} \\
(a) \\
\includegraphics[width=2.4in, angle=0]{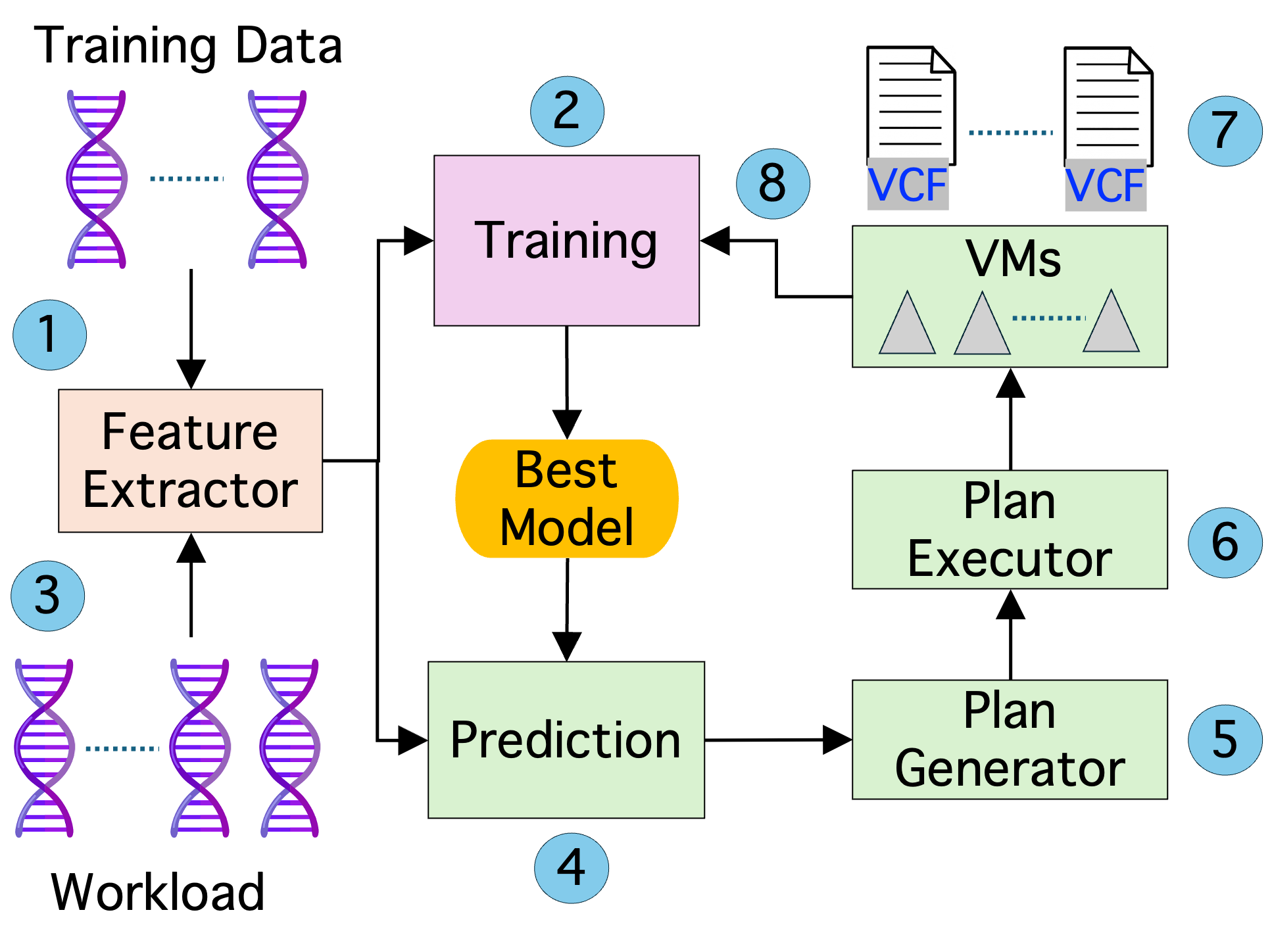} \\
(b) \\
\end{tabular}
\caption{(a) Our computing model~~~~(b) Overview of our approach}
\label{fig-system-model}
\end{center}
\end{figure}

We consider a computing model that can be readily set up in a cloud environment (see Figure~\ref{fig-system-model}(a)). A user provisions a number of VMs, each with different number of GPUs, to execute a variant calling pipeline on a workload of genome sequences. The sequences (in FASTQ format), intermediate files (in BAM format), and output files (in VCF format) are stored in shared storage such as the Network File System (NFS). A stage of the pipeline (for any sequence) is executed on a single VM using all its resources without preemption. (Note that variant calling on distributed GPUs is beyond the scope of this work.)


\subsection{Overview of Our Approach}

Next, we present an overview of our approach (see Figure~\ref{fig-system-model}(b)). To address the challenge of predicting executing times of variant calling pipeline stages, we employ ML. Hence, the first step is to collect training data by executing the variant calling pipeline on available genome sequences and measuring the execution time for different stages on different GPU-enabled VM types. The \textit{Feature Extractor} will extract relevant features from the genome sequences (e.g., using MultiQC\footnote{https://seqera.io/multiqc}). Next, ML models were trained \textit{for different stages and different GPU-enabled VM types} using RF, LR, XGBoost, and NN. Given a new workload of genomes to process, the best ML model type is used to predict the execution time of the variant calling stages on different GPU-enabled VMs. By collecting training data from different GPU-enabled VMs, the ML models can capture the heterogeneity of the execution environment for predicting the execution time of different stages. To address the challenge of formulating the generation of execution plans as an optimization problem, the \textit{Plan Generator} draws inspiration from the FJSP. It also supports a greedy strategy for plan generation. Finally, the \textit{Plan Executor} executes the optimal plans on the VMs by using lightweight synchronization operations to address the final challenge of correctness. This is required because the strict time-based schedules of the Plan Generator may experience variations in execution times of different operations that must be synchronized when executed. VCF files are produced after successful execution of variant calling. The ML models can be retrained when the accuracy of prediction decreases below a threshold based on the actual time taken to process the sequences.

\subsection{ML for Predicting Execution Time of Variant Calling Stages}

We formulate the problem of predicting the execution time of different variant calling stages as \textit{a regression problem}. Our goal is to compute the best model $\Phi$ s.t. $y = \Phi(f_1, \ldots, f_n)$, where \{$f_1, \ldots, f_n$\} denote the feature set based on the characteristics of a sequence, and $y$ is the predicted execution time. Table~\ref{table-features} shows the summary of features used to train the ML models such as sequence size, read quality, \% of duplicate reads, etc. Later, we will demonstrate that these features yield better prediction accuracy compared to solely using the genome size as the feature. For each GPU-enabled VM type and pipeline stage, we train an ML model by measuring execution times on publicly available WGS data on that VM type. For each sequence in the input workload to process, the best ML model is used to estimate the prediction time of the different pipeline stages on different GPU-enabled VMs. This information is then used for optimal execution plan generation. 




\begin{table}[tbh]
\caption{Key Features Extracted for Training ML Models}
\label{table-features}
\begin{center}
\begin{tabular}{|p{1in}|p{2.0in}|}
\hline
\multicolumn{1}{|c|}{\textbf{Feature}} &
\multicolumn{1}{|c|}{\textbf{Description}} \\
\hline
\hline
Size & Size of the genome sequence (in MB) \\
\hline
Avg. length & Average length of reads \\
\hline
Avg. insert size & Length of the DNA fragment between the adapters on each end of a read \\ 
\hline
Spots & Number of sequencing spots (or reads) \\
\hline
Bases & Total number of bases sequenced \\
\hline
Unique reads & Number of unique reads \\
\hline
\% duplicates & \% of duplicate reads \\
\hline
Per base sequence quality & Overall quality score based on the base quality scores \\
\hline
Per base sequence content & Overall \% of A/T/G/C across all the read positions \\
\hline
Per base N content & \% of ambiguous base calls across all each read positions \\
\hline
Per sequence GC content & Measure of the GC content across all reads \\
\hline
Overrepresented reads & Measure of reads that appear more than expected \\
\hline
\end{tabular}
\end{center}
\end{table}


\subsection{Generation of Optimal Execution Plans}

We formulate the problem of generating optimal execution plans for an input workload and VMs as \textit{a combinatorial optimization problem}. Table~\ref{table-notations} shows the list of frequently used notations in the paper.

\begin{table}[tbh]
\caption{Notations and Their Description}
\label{table-notations}
\begin{center}
\begin{tabular}{|p{1in}|p{2.0in}|}
\hline
\multicolumn{1}{|c|}{\textbf{Notation}} &
\multicolumn{1}{|c|}{\textbf{Description}} \\
\hline
\hline
$N$ & Total number of jobs (i.e., genomes) \\
\hline
$M$ & Total number of VMs available \\
\hline
$\mathbb{J} = \{J_1, \ldots, J_N\}$ & Set of jobs to process \\ 
\hline
$J_i = (o_{i1}, \ldots, o_{iK})$ & The sequence of $K$ operations (i.e., stages) of job $J_i$ \\
\hline
$\mathbb{M} = \{m_1, \ldots, m_M\}$ & Set of VMs available to execute the jobs \\
\hline
$T(o_{ij}) = (t_{ij}^{1}, \ldots, t_{ij}^{M})$ & Time taken to execute $o_{ij}$ on the VMs \\ 
\hline
$T(o_{ij}^{k}) = t_{ij}^{k}$ & Time taken by the operation $o_{ij}$ on $m_k$ \\
\hline
$S_i = (\ldots, ( \hat{o}, \hat{s} ), \ldots)$ & Schedule of operations with start times to be executed on $m_i$\\
\hline
$\hat{t}$ & Minimum makespan \\
\hline
$\mathbb{E} = \{e_1, \ldots, e_M\}$ & Set of $M$ execution plans, one per VM \\
\hline
{\tt BEGIN}, {\tt EXEC}, {\tt SIGNAL}, {\tt WAIT}, {\tt END} & Start of a plan; Execute a stage; Signal the next stage to start; Wait for the previous stage to finish; End of a plan \\
\hline
\end{tabular}
\end{center}
\end{table}

We first develop the FJSP-based strategy. It is outlined in Algorithm~\ref{algo-FJSP-strategy}, which is given a set of jobs $\mathbb{J}$ and VMs $\mathbb{M}$, and generates the execution plans $\mathbb{E}$ for the VMs. Each plan is composed of special statements, namely, {\tt BEGIN}, {\tt END}, {\tt EXEC}, {\tt SIGNAL}, and {\tt WAIT}. We assume each $J_i$ can be split into $K$ pipeline stages. Using the predicted execution times for a stage on different VMs (Line~\ref{algo-predicted-time}), the FJSP is solved to obtain optimal schedules (Line~\ref{algo-fjsp}). The optimal schedule $S_j$ for the VM $m_j$ contains the sequence of operations ordered by their start times (Line~\ref{algo-schedule-vm}). For each VM $m_j$, {\tt BEGIN} is appended as the first statement in $e_j$ to indicate the start of the plan (Line~\ref{algo-append-begin}). Next is the plan generation for a VM (see Lines~\ref{algo-plan-generation-begin}-\ref{algo-plan-generation-end}). The first operation of a job can begin immediately; hence, {\tt EXEC} is appended after {\tt BEGIN} to indicate the execution of the operation/stage. For all subsequent operations of a job, {\tt WAIT} is appended to wait for the previous operation/stage of the job to complete for correctness. {\tt EXEC} is appended for every operation. For every operation except the last one of a job, {\tt SIGNAL} is appended to signal the next operation of the job to begin. Hence, {\tt WAIT} and {\tt SIGNAL} provide \textit{lightweight synchronization} across the VMs when the plans are executed to process the workload. Each plan contains {\tt END} to indicate the end of the plan (Line~\ref{algo-append-end}). The synchronization is also needed because the operations can execute slower or faster than the predicted times.



\begin{algorithm}[tbh]
\caption{FJSP-based Strategy for Plan Generation}
\label{algo-FJSP-strategy}
\begin{algorithmic}[1]
\REQUIRE $\mathbb{J}$ - Set of $N$ input jobs; $\mathbb{M}$ - Set of $M$ VMs
\ENSURE \{$e_1$, $e_2$, ..., $e_M$\} - Optimal execution plans
\STATE $\mathbb{J} = \{J_1, J_2, ..., J_N\}$; $\mathbb{M} = \{m_1, m_2, ..., m_M\}$
\STATE Let $J_i = (o_{i1}, ..., o_{iK})$ denote the sequence of $K$ operations of job $j_i$ 
\STATE Let $T(o_{ij}^{k})$ denote the predicted time taken to execute operation $o_{ij}$ on machine $m_k$ using ML \label{algo-predicted-time}
\STATE Generate optimal schedules for jobs $\mathbb{J}$ and VMs $\mathbb{M}$ using the FJSP \label{algo-fjsp}
\STATE Let $\hat{t}$ denote the minimum makespan obtained by the FJSP
\STATE Let $S_{j}=(..., (o_{pq}, s_{pq}), ...)$ denote the schedule of operations ordered by start time for $m_j$ \label{algo-schedule-vm}
\FOR[Generate plans for all machines]{j=1 to M}
\STATE Append {\tt BEGIN} to $e_j$ \COMMENT{First instruction in a plan} \label{algo-append-begin} 
\FOR[Process each operation in a schedule]{each item $(o_{pq}, s_{pq}) \in S_{j}$} \label{algo-plan-generation-begin}
\IF[If not the first operation of a job]{$q > 1$} 
\STATE Append {\tt WAIT}$(o_{pq})$ to $e_j$ \COMMENT{Add WAIT for previous operation to complete}
\ENDIF
\STATE Append {\tt EXEC}$(o_{pq})$ to $e_j$ \COMMENT{Add EXEC for the operation to execute}
\IF[Except the last operation of a job]{$(q+1) \leq K$}
\STATE Append {\tt SIGNAL}$(o_{p(q+1)})$ to $e_j$ \COMMENT{Add SIGNAL for the next operation to start} \label{algo-plan-generation-end}
\ENDIF
\ENDFOR
\STATE Append {\tt END} to $e_j$ \COMMENT{Last instruction in a plan} \label{algo-append-end}
\ENDFOR
\RETURN \{$e_1$, $e_2$, ..., $e_M$\}
\end{algorithmic}
\end{algorithm}

Next, we present the Greedy strategy that may not generate optimal plans but is a good baseline for comparison. Algorithm~\ref{algo-greedy-strategy} outlines the steps involved. Each plan has {\tt BEGIN} as the first statement (Line~\ref{init-plan-end}). Next, the total predicted time for each job on a VM is computed after applying ML (Line~\ref{algo-init-job-time-1}-\ref{algo-init-job-time-2}). For each unassigned job, we find the VM that requires the least amount of time (Lines~\ref{algo-fastest-VM-begin}-\ref{algo-fastest-VM-end}). From these selected job/VM pairs, we pick the job/VM pair that requires the least amount of time and assign that job to that machine (Line~\ref{algo-min-job-VM-pair}). {\tt EXEC} statements are appended to the plan of that machine to execute all the operations of that job (Lines~\ref{algo-exec-for-a-job-begin}-\ref{algo-exec-for-a-job-end}). The remaining jobs are assigned similarly to the remaining machines until all the machines are assigned at least one job. The steps are repeated until all jobs are assigned. As each job is entirely executed on a single VM, there are no {\tt WAIT}/{\tt SIGNAL} statements. {\tt END} is appended to all the plans (Line~\ref{algo-end-greedy-plan}).

\begin{algorithm}[h]
\caption{Greedy Strategy for Plan Generation}
\label{algo-greedy-strategy}
\begin{algorithmic}[1]
\REQUIRE $\mathbb{J}$ - Set of $N$ input jobs; $\mathbb{J}$ - Set of $M$ VMs
\ENSURE \{$e_1$, $e_2$, ..., $e_M$\} - Execution plan for the machines
\FOR{$j=1$ to M} \label{init-plan-begin} 
\STATE Append {\tt BEGIN} to $e_j$ \label{init-plan-end} \COMMENT{First instruction in a plan}
\ENDFOR
\FOR{$i=1$ to N} \label{algo-init-job-time-1}
\FOR{$j=1$ to M}
\STATE Let $T(o_{ij}^{k})$ denote the predicted time for $o_{ij}$ on $m_k$ using ML
\STATE $W_{ij} = \sum_{r=1}^{K} T(o_{ir}^{j})$ \label{algo-init-job-time-2} \COMMENT{Compute total predicted time for a job on a machine}
\ENDFOR
\ENDFOR
\STATE $\widehat{\mathbb{J}} \leftarrow \mathbb{J}$ \COMMENT{Initialize jobs to be assigned}
\FOR[Iterate through each job]{$i=1$ to N}
\STATE $\widehat{\mathbb{M}} \leftarrow \mathbb{M}$ \COMMENT{Initialize available machines}
\FOR[Iterate through each machine]{$j=1$ to M}
\STATE $\mathbb{T} = \emptyset$
\FOR[Iterate through unassigned jobs]{each $J_p \in \widehat{\mathbb{J}}$} \label{algo-fastest-VM-begin}
\STATE $q \leftarrow \argminl_{j~s.t.~m_j \in \widehat{\mathbb{M}}}(W_{pj})$ \COMMENT{Find the fastest machine for the job} 
\STATE $\mathbb{T} \leftarrow \mathbb{T} \cup \{(J_p, m_q, W_{pq})\}$ \label{algo-fastest-VM-end}
\ENDFOR
\STATE Find $(J_{\alpha}, m_{\beta}, w) \in \mathbb{T}$ s.t. $w$ is the minimum time  \label{algo-min-job-VM-pair}
\STATE \COMMENT{Find the job/machine pair that requires the least time}
\FOR{$k=1$ to K} \label{algo-exec-for-a-job-begin}
\STATE Append {\tt EXEC}$(o_{\alpha{}k})$ to $e_{\beta}$ \COMMENT{Add {\tt EXEC} for each operation of the job} \label{algo-exec-for-a-job-end}
\ENDFOR
\STATE $\widehat{\mathbb{J}} \leftarrow \widehat{\mathbb{J}} \setminus \{J_{\alpha}\}$ \COMMENT{Remove the assigned job from further consideration}
\STATE $\widehat{\mathbb{M}} \leftarrow \widehat{\mathbb{M}} \setminus \{m_{\beta}\}$ \COMMENT{Remove the assigned machine from further consideration}
\ENDFOR
\ENDFOR
\FOR{$j=1$ to M} 
\STATE Append {\tt END} to $e_j$ \COMMENT{Last instruction in a plan} \label{algo-end-greedy-plan}
\ENDFOR
\RETURN \{$e_1$, $e_2$, ..., $e_M$\}
\end{algorithmic}
\end{algorithm}

We provide an example to illustrate the differences between the FJSP-based and Greedy strategies.

\begin{example}
\label{job-shop-example1}
Consider three jobs $J_1$, $J_2$, and $J_3$ with three operations each. Let $m_1$, $m_2$, and $m_3$ denote the VMs. Let $J_1 = (o_{11}, o_{12}, o_{13})$, $J_2 = (o_{21}, o_{22}, o_{23})$, and $J_3 = (o_{31}, o_{32}, o_{33})$. Let $T(o_{11}) = (3, 2, 5)$, $T(o_{12}) = (2, 4, 4)$, and $T(o_{13}) = (4, 3, 1)$ denote the time taken to run an operation of $J_1$ on each VM. Similarly, let $T(o_{21}) = (3, 3, 4)$, $T(o_{22}) = (1, 5, 3)$, and $T(o_{23}) = (2, 2, 5)$ denote the execution times of the operations for $J_2$; and let $T(o_{31}) = (3, 2, 5)$, $T(o_{32}) = (5, 3, 3)$, and $T(o_{33}) = (3, 2, 4)$. \hfill $\Box$
\end{example}

\begin{figure}[tbh]
\begin{center}
\begin{tabular}{c}
\includegraphics[width=1.8in, angle=0]{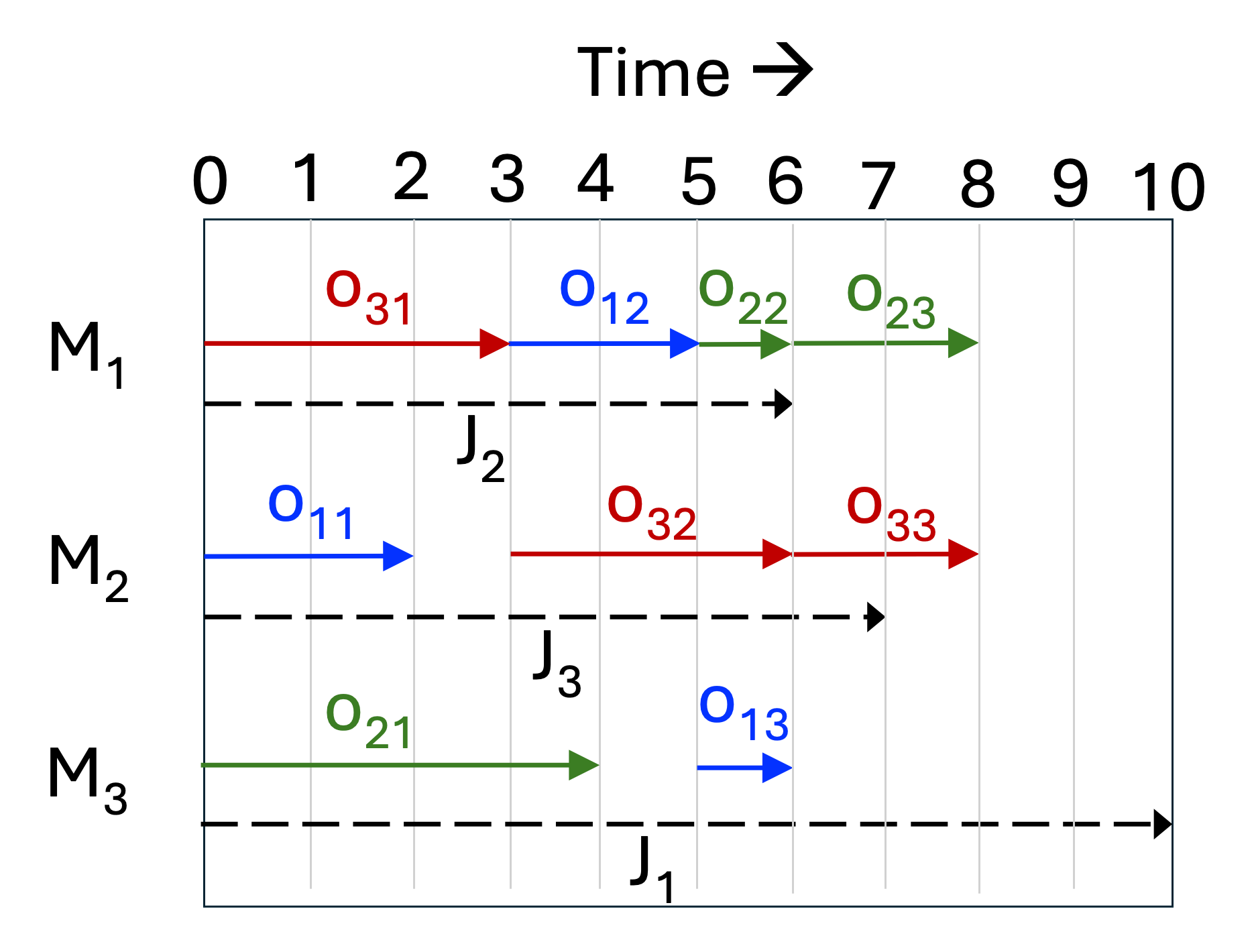} \\
(a) Generated schedules \\
\includegraphics[width=1.8in, angle=0]{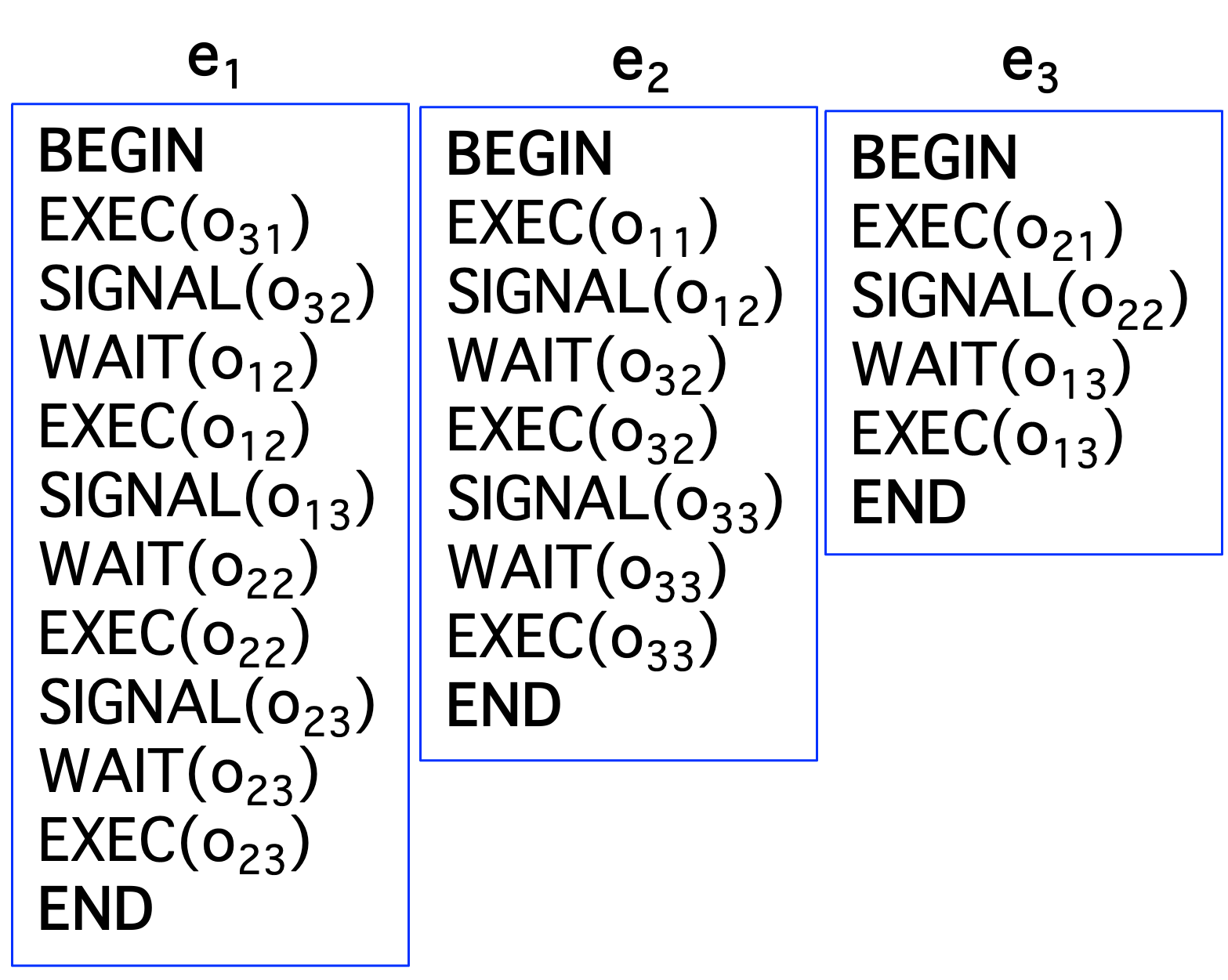} \\
(b) FJSP-based execution plan \\
\includegraphics[width=1.65in, angle=0]{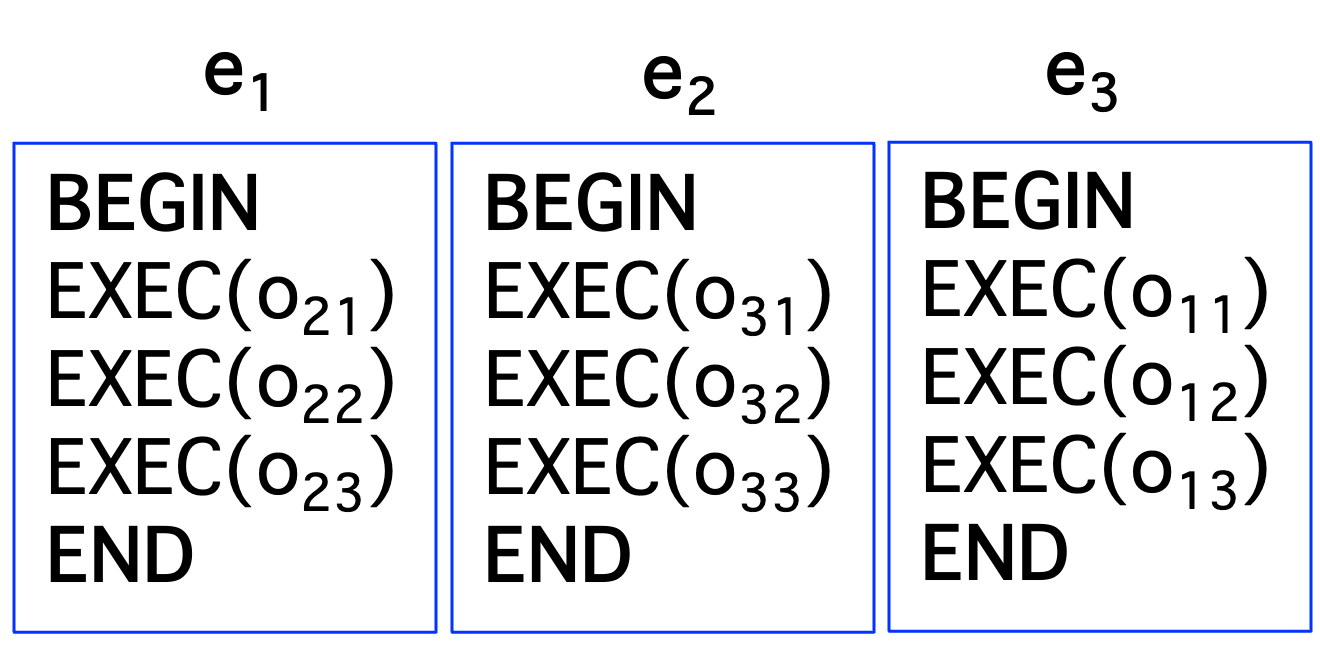} \\
(c) Greedy execution plan
\end{tabular}
\caption{Schedules/Execution Plans: FJSP-based strategy vs. Greedy strategy}
\label{fig-example-schedule}
\end{center}
\end{figure}

\begin{example}
\label{job-shop-example2}
Based on the FJSP-based strategy (Algorithm~\ref{algo-FJSP-strategy}), we obtain optimal execution plans for each VM with the makespan of 8. The schedules are shown in Figure~\ref{fig-example-schedule}(a) by colored lines. For example, $S_1 = ((o_{31}, 0), (o_{12}, 3), (o_{22}, 5), (o_{23}, 6))$. As observed, each stage of a job can be executed by a different VM. The corresponding execution plans are shown in Figure~\ref{fig-example-schedule}(b). \hfill $\Box$
\end{example}



\begin{example}
\label{greedy-example3}
Based on the Greedy strategy (Algorithm~\ref{algo-greedy-strategy}), $J_2$ is first assigned to $M_1$ ($w=6$), $J_3$ is then assigned to $M_2$ ($w=7$), and finally, $J_1$ is assigned to $M_3$ ($w=10$). The makespan is 10 as shown by dotted lines in Figure~\ref{fig-example-schedule}(a). The corresponding execution plans are shown in Figure~\ref{fig-example-schedule}(c). \hfill $\Box$
\end{example}



\subsection{Plan Execution on the VMs}

We discuss how the Plan Executor processes the execution plans on the VMs while ensuring correctness. Algorithm~\ref{algo-plan-execution} outlines the steps involved and considers plans for both the Greedy and FJSP-based strategies. The Plan Executor initializes the start time when {\tt BEGIN} is encountered. It blocks to obtain a specific file lock when {\tt WAIT} is encountered. Once the file lock is obtained, the {\tt EXEC} statement is processed by invoking a variant calling stage. After that, when {\tt SIGNAL} is encountered, a new file is created on shared storage to signal the completion of the current operation. The end time is recorded when {\tt END} is processed. Timeouts can be added for file locking in case certain stages fail during execution. 


\begin{algorithm}[tbh]
\caption{Plan Execution}
\label{algo-plan-execution}
\begin{algorithmic}[1]
\REQUIRE $e_i$ - Execution plan for machine $m_i$
\ENSURE Makespan
\FOR[Iterate through each instruction]{each $op(param) \in e_i$} 
\IF{$op$ $=$ {\tt BEGIN}}
\STATE $beginTime \leftarrow gettime()$ \COMMENT{Record start time}
\ELSIF{$op$ $=$ {\tt WAIT}}
\STATE {\tt DOWAIT}($param$) \COMMENT{Wait for previous operation to complete}
\ELSIF{$op$ $=$ {\tt EXEC}} 
\STATE Run variant calling pipeline stage specified in $param$
\ELSIF{$op$ $=$ {\tt SIGNAL}}
\STATE {\tt DOSIGNAL}($param$) \COMMENT{Signal the next operation to execute}
\ELSIF{$op$ $=$ {\tt END}}
\STATE $endTime \leftarrow gettime()$ \COMMENT{Record end time}
\ENDIF
\ENDFOR
\RETURN $endTime - beginTime$
\end{algorithmic}
\end{algorithm}

\subsection{Dynamic Strategy}

To compare against the earlier approaches that generate static schedules using ML-based time predictions, we developed a dynamic scheduling strategy based on the master-worker model. The master maintains the list of sequences to process. It periodically checks the worker VMs if they are free or busy processing sequences. If a worker VM is free, the master assigns the next sequence in the list to that VM. The master continues until all the sequences are assigned and processed. Each sequence is processed entirely by only one VM using all its GPUs. This strategy is dynamic and does not use ML-based time predictions. Algorithm~\ref{algo-Dynamic-strategy} outlines the steps involved in the Dynamic strategy. 

\begin{algorithm}[tbh]
\caption{Dynamic Strategy (using the master-worker model)}
\label{algo-Dynamic-strategy}
\begin{algorithmic}[1]
\REQUIRE $\mathbb{J}$ - Set of $N$ input jobs; $\mathbb{M}$ - Set of $M$ VMs
\ENSURE Makespan 
\STATE $\mathbb{J} = \{J_1, J_2, ..., J_N\}$; $\mathbb{M} = \{m_1, m_2, ..., m_M\}$
\STATE Let $s$ denote time (e.g., 30 s)
\STATE $beginTime \leftarrow gettime()$ \COMMENT{Record start time}
\FOR[Iterate through the job list]{$j=1$ to $N$}
\WHILE{True}
\STATE $f \leftarrow 0$
\FOR[Iterate through the list of VMs]{$i=1$ to $M$}
\IF{$m_i$ is free} 
\STATE $f \leftarrow i$; break \COMMENT{Found a free VM}
\ENDIF
\ENDFOR
\IF{$f \neq 0$}
\STATE break 
\ELSE
\STATE sleep($s$) \COMMENT{No free VMs; so wait and check again}
\ENDIF
\ENDWHILE
\STATE Assign $J_j$ to $m_{f}$ for execution of the variant calling pipeline
\ENDFOR
\STATE Wait for all jobs to finish \COMMENT{The master must wait for all workers to finish}
\STATE $endTime \leftarrow gettime()$ \COMMENT{Record end time}
\RETURN $endTime - beginTime$
\end{algorithmic}
\end{algorithm}

\section{Performance Evaluation}
\label{sec-evaluation}

In this section, we compare the performance of the FJSP-based, Greedy, and Dynamic strategies for optimized execution of variant calling pipelines on human genomes. Recall that the ML-based time predictions are used only by the FJSP-based and Greedy strategies.

\subsection{Experimental Setup}
We conducted all our experiments on FABRIC~\cite{FABRIC}, a unique international research infrastructure for networking/distributed computing research and development of new science applications. We allocated 5 VMs with a total of 9 GPUs attached to them. (The physical servers had AMD EPYC 7532 32-core processors.) Each VM had 24 virtual CPUs, 64 GB RAM, and 1TB storage. They ran Ubuntu Linux (22.04 LTS). The hardware configuration of the VMs are shown in Table~\ref{table-hardware-configuration}. One VM ran the NFS server; the other VMs ran the NFS clients. All GPUs of a VM were used to run the variant calling pipelines.



\begin{table}
\caption{Hardware configuration of the VMs}
\label{table-hardware-configuration}
\begin{center}
\begin{tabular}{|c|c|c|c|c|c|}
\hline
\multicolumn{1}{|c|}{VM} &
\multicolumn{1}{|c|}{Virtual} &
\multicolumn{1}{|c|}{RAM} &
\multicolumn{1}{|c|}{Storage} &
\multicolumn{1}{|c|}{GPU Count} &
\multicolumn{1}{|c|}{Total GPU} \\
\multicolumn{1}{|c|}{ID} &
\multicolumn{1}{|c|}{CPUs} &
\multicolumn{1}{|c|}{} &
\multicolumn{1}{|c|}{} &
\multicolumn{1}{|c|}{and Type} &
\multicolumn{1}{|c|}{VRAM} \\
\hline
\hline
$vm_1$ & 24 & 64 GB & 1 TB & 2 Tesla T4 & 32 GB \\
\hline
$vm_2$ & 24 & 64 GB & 1 TB & 1 Tesla T4 & 16 GB \\
\hline
$vm_3$ & 24 & 64 GB & 1 TB & 3 RTX 6000 & 76 GB \\
\hline
$vm_4$ & 24 & 64 GB & 1 TB & 2 RTX 6000 & 48 GB \\
\hline
$vm_5$ & 24 & 64 GB & 1 TB & 1 RTX 6000 & 24 GB \\
\hline
\end{tabular}
\end{center}
\end{table}



Our code was developed in Python (version 3.12.8). To solve the FJSP to generate optimal execution plans, we used Google OR-Tools\footnote{https://developers.google.com/optimization} that employs the CP-SAT solver~\cite{Perron2023}. We used NVIDIA Parabricks 4.1.0~\cite{OConnell2022}, a GPU-optimized open-source software for variant calling. Parabricks runs on a single machine but can leverage multiple GPUs on the machine for parallelism. We used the \textit{low memory} option in Parabricks as it could not process some of the sequences on the tested GPUs. The CUDA 12.0 software was used. The germline variant calling pipeline of Parabricks reads FASTQ files and performs alignment using BWA-MEM~\cite{BWAMEM2013}. It then performs sorting of the mapped reads and marking of duplicates followed by BQSR. Finally, HaplotypeCaller~\cite{HaplotypeCaller} is invoked to generate the VCF. We executed Parabricks in two modes: 1-stage (FASTQ$\rightarrow$VCF) and 2-stage (FASTQ$\rightarrow$BAM and BAM$\rightarrow$VCF). This enables us to evaluate if smaller operations are better for the FJSP-based strategy.

\subsection{Dataset and Metrics}

We used 98 publicly available whole genome sequences from the 1000 Genomes Project\footnote{https://www.internationalgenome.org}. The total size of these low-coverage (paired-end) sequences was 632 GB (in compressed form). The min. and max. size of the sequences (in compressed form) were 2.2 GB and 15.4 GB, respectively.


We measured regression metrics, namely, $R^2$, Mean Squared Error (MSE), and Mean Absolute Error (MAE) for evaluating the models for predicting execution times. (As we trained a separate ML model for each VM type, we report the average values of the metrics.) We also compared the predicted makespan with the actual makespan for each strategy. Our goal was to show that the FJSP-based strategy can achieve faster execution of a workload compared to the Greedy and Dynamic strategies. We also measured the CPU and GPU utilization of the VMs for different strategies.

\subsection{Results}

\begin{table}[h]
\caption{Performance of ML Models for Predicting Execution Times (in Seconds) Averaged Across All VMs Types (Best Average $R^2$ Values Shown in Bold)}
\label{table-ML-performance-1-stage}
\begin{center}
\begin{tabular}{|p{0.4in}|c|c|c|c|c|c|}
\hline
\multicolumn{1}{|c|}{} &
\multicolumn{6}{|c|}{FASTQ$\rightarrow$VCF} \\
\cline{2-7}
\multicolumn{1}{|c|}{Model} &
\multicolumn{3}{|c|}{One Feature (Size Only)} &
\multicolumn{3}{|c|}{All Features (Incl. Size)} \\
\cline{2-7}
\multicolumn{1}{|c|}{} &
\multicolumn{1}{|c|}{R$^2$} &
\multicolumn{1}{|c|}{MSE} &
\multicolumn{1}{|c|}{MAE} &
\multicolumn{1}{|c|}{R$^2$} &
\multicolumn{1}{|c|}{MSE} &
\multicolumn{1}{|c|}{MAE} \\
\hline
\hline
RF & 0.718 &  10.13 & 192.14 & \textbf{0.894} & 5.43 & 65.09 \\
\hline
XGBoost & 0.628 & 11.18 & 233.27 & 0.890 & 5.40 & 64.94 \\
\hline
LR & 0.674 & 11.11 & 219.27 & 0.732 & 10.11 & 175.76 \\
\hline
\end{tabular}
\end{center}
\end{table}

\begin{table}[h]
\caption{Performance of ML Models for Predicting Execution Times (in Seconds) Averaged Across All VM Types (Best Average $R^2$ Values Shown in Bold)}
\label{table-ML-performance-2-stage}
\begin{center}
\begin{tabular}{|c|c|c|c|c|c|c|}
\cline{2-7}
\multicolumn{1}{c}{} &
\multicolumn{6}{|c|}{FASTQ$\rightarrow$BAM Stage} \\
\cline{1-7}
\multicolumn{1}{|c|}{} &
\multicolumn{3}{|c|}{One Feature} &
\multicolumn{3}{|c|}{All Features} \\
\multicolumn{1}{|c|}{\textbf{Model}} &
\multicolumn{3}{|c|}{(Size Only)} &
\multicolumn{3}{|c|}{(Including Size)} \\
\cline{2-7}
\multicolumn{1}{|c|}{} &
\multicolumn{1}{|c|}{R$^2$} &
\multicolumn{1}{|c|}{MSE} &
\multicolumn{1}{|c|}{MAE} &
\multicolumn{1}{|c|}{R$^2$} &
\multicolumn{1}{|c|}{MSE} &
\multicolumn{1}{|c|}{MAE} \\
\hline
\hline
RF & 0.80 & 7.83 & 116.02 & \textbf{0.904} & 5.54 & 56.40 \\ 											
\hline
XGBoost & 0.73 & 9.29 & 158.59 & 0.874 & 6.32 & 73.10 \\ 															
\hline
LR & 0.72 & 10.26 & 160.26 & 0.772 & 9.69 & 133.47  \\ 													
\hline
\multicolumn{1}{c}{} & \multicolumn{6}{|c|}{BAM$\rightarrow$VCF Stage} \\
\hline
RF & 0.65 & 10.11 & 169.21 & \textbf{0.804} & 6.30 & 94.10 \\ 											
\hline
XGBoost & 0.41 & 12.42 & 274.80 & 0.774 & 6.85 & 110.52 \\ 															
\hline
LR & 0.55 & 11.5 & 223.64 & 0.664 & 9.87 & 166.43 \\ 													
\hline
\end{tabular}
\end{center}
\end{table}

\begin{table*}[tbh]
\caption{Performance Comparison: FJSP-based vs. Greedy Strategy vs. Dynamic Strategy (best makespan in bold)}
\label{table-greedy-FJSP-dynamic-comparison}
\begin{center}
\begin{tabular}{|c|r|r|r|r|c|c|c|}
\multicolumn{7}{c}{\# of machines (\textbf{M}): 5 \hspace{5em} \# of sequences per subset (\textbf{N}): 10} \\   
\hline
\multicolumn{1}{|c|}{} & 
\multicolumn{4}{|c|}{Makespan} &
\multicolumn{1}{|c|}{Best Speedup} &
\multicolumn{1}{|c|}{Best Speedup} \\
\cline{2-5}
\multicolumn{1}{|c|}{Subset} & 
\multicolumn{1}{|c|}{Greedy} & 
\multicolumn{1}{|c|}{Dynamic} &
\multicolumn{1}{|c|}{FJSP} &
\multicolumn{1}{|c|}{FJSP} &
\multicolumn{1}{|c|}{of FJSP } &
\multicolumn{1}{|c|}{of FJSP } \\
\multicolumn{1}{|c|}{ID} & 
\multicolumn{1}{|c|}{Strategy} & 
\multicolumn{1}{|c|}{Strategy} &
\multicolumn{1}{|c|}{Strategy} &
\multicolumn{1}{|c|}{Strategy} &
\multicolumn{1}{|c|}{(2-stage)} & 
\multicolumn{1}{|c|}{(2-stage)} \\
\multicolumn{1}{|c|}{} & 
\multicolumn{1}{|c|}{} & 
\multicolumn{1}{|c|}{} &
\multicolumn{1}{|c|}{(1-stage)} &
\multicolumn{1}{|c|}{(2-stage)} &
\multicolumn{1}{|c|}{w.r.t. Greedy} & 
\multicolumn{1}{|c|}{w.r.t. Dynamic} \\
\hline
\hline
1 & 9,729 s & 6,362 s & 6,730 s & \textbf{5,104 s} & 1.90$\times$ & 1.24$\times$\\
\hline
2 & 10,628 s & 8,001 s & 7,226 s & \textbf{5,118 s} &  2.07$\times$ & 1.56$\times$\\
\hline
3 & 10,498 s & 8,524 s & 6,967 s & \textbf{5,687 s} &  1.84$\times$ & 1.49$\times$\\
\hline
4 & 10,491 s & 6,984 s & 5,972 s & \textbf{4,922 s} & 2.13$\times$ & 1.41$\times$\\
\hline
5 & 10,536 s & 9,735 s & 5,961 s & \textbf{4,703 s} & 2.24$\times$ & 2.06$\times$\\
\hline
6 & 10,457 s & 10,575 s & 6,744 s & \textbf{5,618 s} & 1.86$\times$ & 1.88$\times$\\
\hline
7 & 10,343 s & 8,559 s & 5,575 s & \textbf{4,151 s} & 2.49$\times$ & 2.06$\times$\\
\hline
8 & 10,317 s & 9,059 s & 6,562 s & \textbf{5,782 s} & 1.78$\times$ & 1.56$\times$\\
\hline
9 & 10,703 s & 8,051 s & 7,670 s & \textbf{6,346 s} & 1.68$\times$ & 1.26$\times$\\
\hline
\hline
\textit{Average} & 10,411.3 s & 8,427.7 s & 6,600.7 s & 5,270.1 s & 2.00$\times$ & 1.61$\times$ \\
\hline
\end{tabular}
\end{center}
\end{table*}

\begin{table*}[tbh]
\caption{Effectiveness of ML (RF model) in Predicting Makespan}
\label{table-predicted-comparison}
\begin{center}
\begin{tabular}{|c|r|r|r|r|r|r|r|r|r|}
\multicolumn{10}{c}{\# of machines (\textbf{M}): 5 \hspace{5em} \# of sequences per subset (\textbf{N}): 10} \\   
\hline
\multicolumn{1}{|c|}{Subset} & 
\multicolumn{3}{|c|}{Greedy} & 
\multicolumn{3}{|c|}{FJSP (1-stage)} &
\multicolumn{3}{|c|}{FJSP (2-stage)} \\
\cline{2-10}
\multicolumn{1}{|c|}{ID} & 
\multicolumn{1}{|c|}{Predicted} & 
\multicolumn{1}{|c|}{Actual} &
\multicolumn{1}{|c|}{RE} &
\multicolumn{1}{|c|}{Predicted} &
\multicolumn{1}{|c|}{Actual} &
\multicolumn{1}{|c|}{RE} &
\multicolumn{1}{|c|}{Predicted} &
\multicolumn{1}{|c|}{Actual} &
\multicolumn{1}{|c|}{RE} \\
\multicolumn{1}{|c|}{} & 
\multicolumn{1}{|c|}{Makespan} & 
\multicolumn{1}{|c|}{Makespan} &
\multicolumn{1}{|c|}{(\%)} &
\multicolumn{1}{|c|}{Makespan} &
\multicolumn{1}{|c|}{Makespan} &
\multicolumn{1}{|c|}{(\%)} &
\multicolumn{1}{|c|}{Makespan} &
\multicolumn{1}{|c|}{Makespan} &
\multicolumn{1}{|c|}{(\%)} \\
\hline
\hline
1&	9,860 s&	9,729 s&	1.34&	5,457 s&	6,730 s&	18.91&	5,259 s&	5,104 s&	3.03 \\
\hline
2&	9,927 s&	10,628 s&	6.59&	5,857 s&	7,226 s&	18.94&	5,575 s&	5,118 s&	8.92 \\
\hline
3&	10,224 s&	10,498 s&	2.61&	6,237 s&	6,967 s&	10.47&	4,811 s&	5,687 s&	15.40 \\
\hline
4&	9,952 s&	10,491 s&	5.13&	5,497 s&	5,972 s&	7.95&	4,817 s&	4,922 s&	2.13 \\
\hline
5&	9,745 s&	10,536 s&	7.50&	5,450 s&	5,961 s&	8.57&	5,220 s&	4,703 s&	10.99 \\
\hline
6&	9,952 s&	10,457 s&	4.82&	5,533 s&	6,744 s&	17.95&	4,830 s&	5,618 s&	14.02 \\
\hline
7&	9,745 s&	10,343 s&	5.78&	4,690 s&	5,575 s&	15.87&	5,776 s&	4,151 s&	39.14 \\
\hline
8&	9,745 s&	10,317 s&	5.54&	5,533 s&	6,562 s&	15.68&	4,811 s&	5,782 s&	16.79 \\
\hline
9&	10,133 s&	10,703 s&	5.32&	6,298 s&	7,670 s&	17.88&	5,816 s&	6,346 s&	8.35 \\
\hline
\hline
\multicolumn{3}{|c|}{\textit{Average RE} (\%)} &  4.96 & \multicolumn{2}{|c|}{\textit{Average RE} (\%)} & 14.69 & \multicolumn{2}{|c|}{\textit{Average RE} (\%)} & 13.20 \\
\hline
\end{tabular}
\end{center}
\end{table*}

\paragraph{Predicting Execution Times of Variant Calling Stages.}

First, we report the performance of ML models for predicting the execution time of the variant calling pipeline. ML models based on LR, RF, XGBoost, SVM, and NN were used to train and test using 80 sequences. (The remaining 18 sequences were used to evaluate the FJSP-based and Greedy strategies.) For each \textit{GPU-enabled VM type} and \textit{pipeline stage}, a separate ML model was trained. We considered two scenarios for building the models: The first scenario used only the size of genome as the feature. The second scenario used all the features shown in Table~\ref{table-features}, which included the genome size. Table~\ref{table-ML-performance-1-stage} and Table~\ref{table-ML-performance-2-stage} show the ML model performance with 10-fold cross validation on the 80 sequences for 1-stage and 2-stage execution, respectively. Note that the reported regression metrics denote the average across the different GPU-enabled VM types. (SVM and NN performed worse than the others and are not reported in the table.)  RF was the best model in both cases. For example, RF achieved an $R^2$ score of \textbf{0.894} for \textit{1-stage} execution. It achieved a score of \textbf{0.904} (FASTQ$\rightarrow$BAM) and \textbf{0.804} (BAM$\rightarrow$VCF) for \textit{2-stage} execution. Further, models that used all the features performed significantly better than just using the genome size. Thus, the \textit{characteristics of the genome sequences} (e.g., size, read quality, \% duplicates, per sequence GC content) impact the execution speed of variant calling.


\paragraph{FJSP-based Strategy vs. Greedy Strategy vs. Dynamic Strategy}



Next, we compare the execution plans of the FJSP-based strategy and the Greedy strategy. Using 18 new genome sequences that were not considered during model training, we randomly generated 9 subsets containing 10 sequences each. Each subset was executed using the FJSP-based strategy (1-stage and 2-stage), the Greedy strategy, and the Dynamic strategy. RF was used to predict execution times for the FJSP-based and Greedy strategies. Table~\ref{table-greedy-FJSP-dynamic-comparison} shows the makespan of each strategy on each subset. Across all the tested subsets, the Greedy strategy had the worst makespan, and the FJSP-based strategy (2-stage) had the best makespan. While the Dynamic strategy was better than the Greedy strategy, it failed to outperform the FJSP-based strategies in most cases. Clearly, this demonstrates that generating optimized static schedules using the FJSP and ML-based execution time predictions \textit{yields better performance} than a dynamic schedule.

Table~\ref{table-greedy-FJSP-dynamic-comparison} also reports the best speedup achieved by the FJSP-based strategy (2-stage) compared to the Greedy and Dynamic strategies. It achieved \textbf{$2\times$} speedup (on an average) over the Greedy strategy and \textbf{$1.61\times$} speedup (on an average) over the Dynamic strategy. While the \textit{2-stage} FJSP strategy required separate ML models for predicting execution times, it enabled superior execution plans compared to the \textit{1-stage} FJSP strategy due to shorter tasks/operations.

To understand the effectiveness of ML for generating optimized execution plans, we compared the predicted and actual makespans for the Greedy and FJSP strategies on the tested subsets. Recall that RF was used for predicting the execution times. The makespan and the average relative error (RE) values are reported in Table~\ref{table-predicted-comparison}. Using RF, the average RE was under \textbf{15\%}. This shows that ML was effective in predicting the execution times of variant calling pipeline stages. This enabled the FJSP-based strategies to generate optimized execution plans leading to faster execution of the tested workload.

\subsection{Resource Utilization}

To gain better understanding of the different strategies, we also measured the CPU and GPU utilization of the 5 VMs during execution of the subsets. Note that we used a heterogeneous execution environment for the experiments as reported in Table~\ref{table-hardware-configuration}. For instance, \textit{vm2} had the least GPU computing power, and \textit{vm3} had the best GPU computing power. \textit{Appendix A} reports the CPU/GPU plots for a representative subset. Figure~\ref{fig-cpu-utilization-plots} shows the CPU utilization of the VMs in terms of 1-min load average (measured in 30-second intervals using \textit{dstat}) for the representative subset. Figure~\ref{fig-gpu-utilization-plots} shows the GPU utilization of the VMs (measured using \textit{nvidia-smi}) for the same subset. As observed, the FJSP-based strategies led to more balanced usage of VM resources in the heterogeneous environment compared to the Greedy and Dynamic strategies. Hence, they achieved \textit{better average makespan} on the tested subsets.  







\section{Conclusion}
\label{sec-conclusion}

We developed a new ML-based approach for optimizing the execution of variant calling pipelines on a genome workload using GPU-enabled VMs. Our approach employed ML to predict the execution times of different variant calling pipeline stages by using different characteristics of a genome sequence. Using the predicted times, our approach generated optimal execution plans by solving the FJSP and then executed them via careful synchronization across the given VMs. Using publicly available WGS data, we showed the effectiveness of ML for predicting execution times. Our FJSP-based strategy achieved $2\times$ speedup (on an average) over the Greedy strategy. It also achieved $1.6\times$ speedup (on an average) over the Dynamic strategy that did not use any ML-based time predictions. Hence, our approach can provide substantial speedup and cost savings for processing human genomes especially in a cloud environment with the \textit{pay-as-you-go} pricing model. 


\begin{acks}
This work was supported by the National Science Foundation under Grant No. 2201583.
\end{acks}

\bibliographystyle{ACM-reference-format}
\bibliography{bigdata}
\appendix
\paragraph{\textbf{Appendix A}}




\begin{figure*}[tbh!]
\begin{tabular}{cccc}
\hspace{-1em}
\includegraphics[width=1.62in, angle=0]{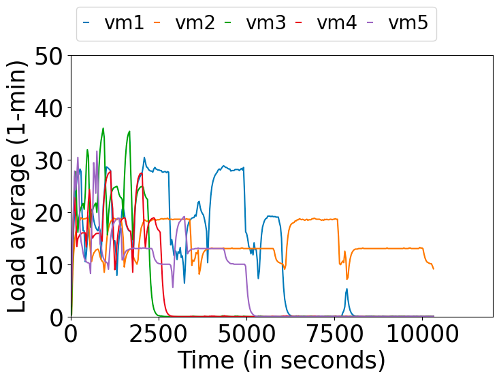} &
\hspace{-1em}
\includegraphics[width=1.7in, angle=0]{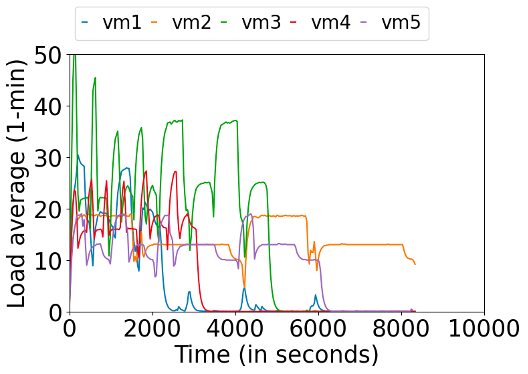} & 
\hspace{-1em}
\includegraphics[width=1.7in, angle=0]{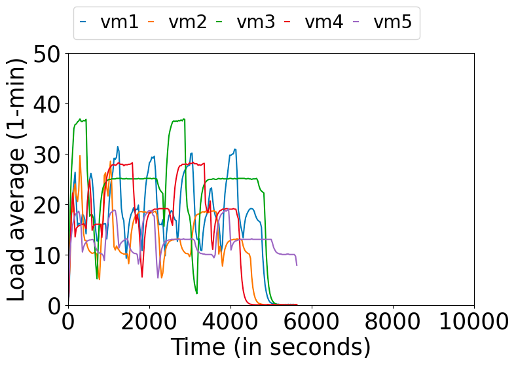} &
\hspace{-1em}
\includegraphics[width=1.62in, angle=0]{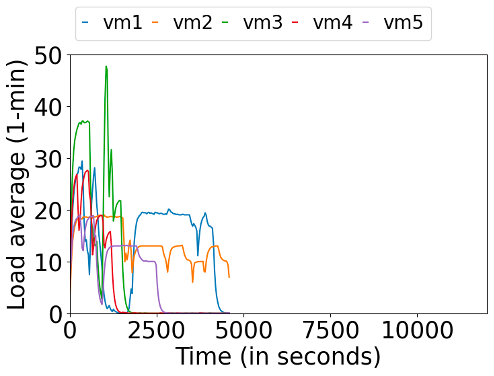} \\
(a) Greedy Strategy & (b) Dynamic Strategy &
(c) FJSP Strategy (1-stage) & (d) FJSP Strategy (2-stage) \\
\end{tabular}
\caption{CPU utilization plots (for the 5 VMs) for different strategies on a representative subset}
\label{fig-cpu-utilization-plots}
\end{figure*}

\begin{figure*}[tbh!]
\begin{tabular}{cccc}
\includegraphics[width=1.70in, angle=0]{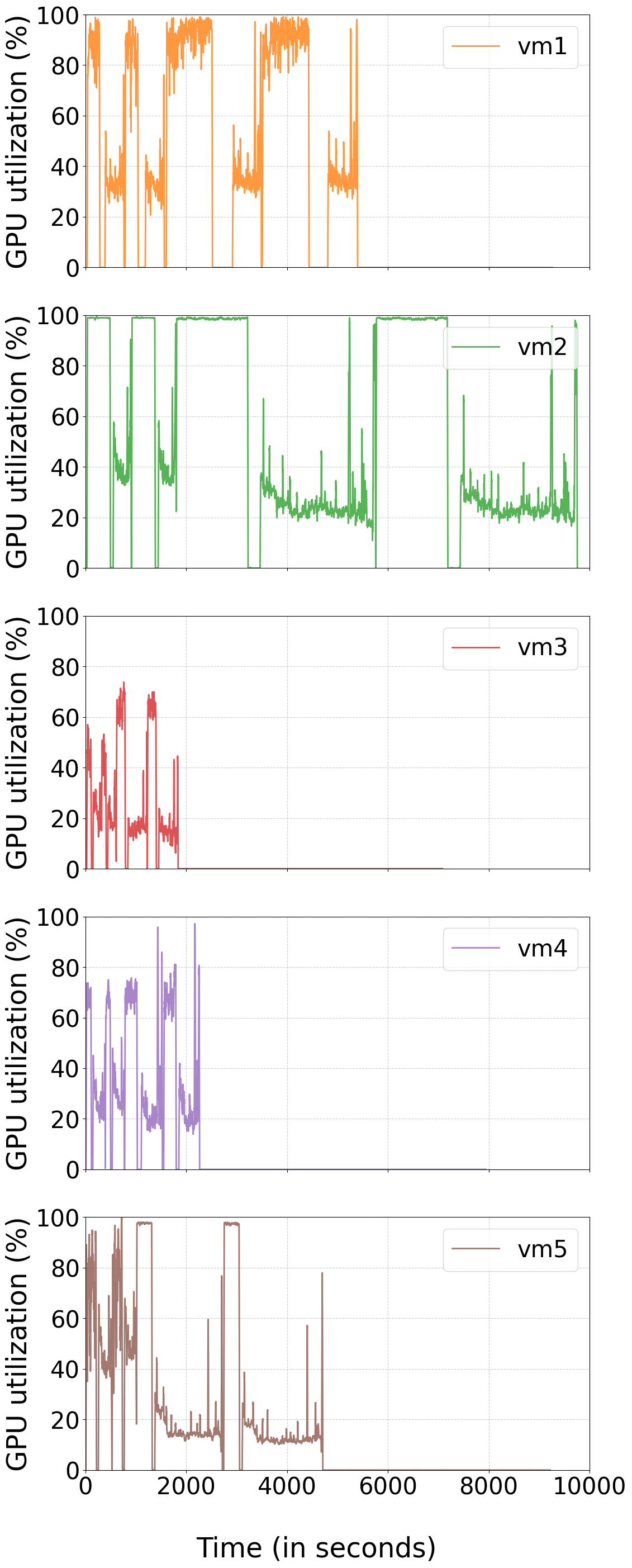} &
\hspace{-1em}
\includegraphics[width=1.70in, angle=0]{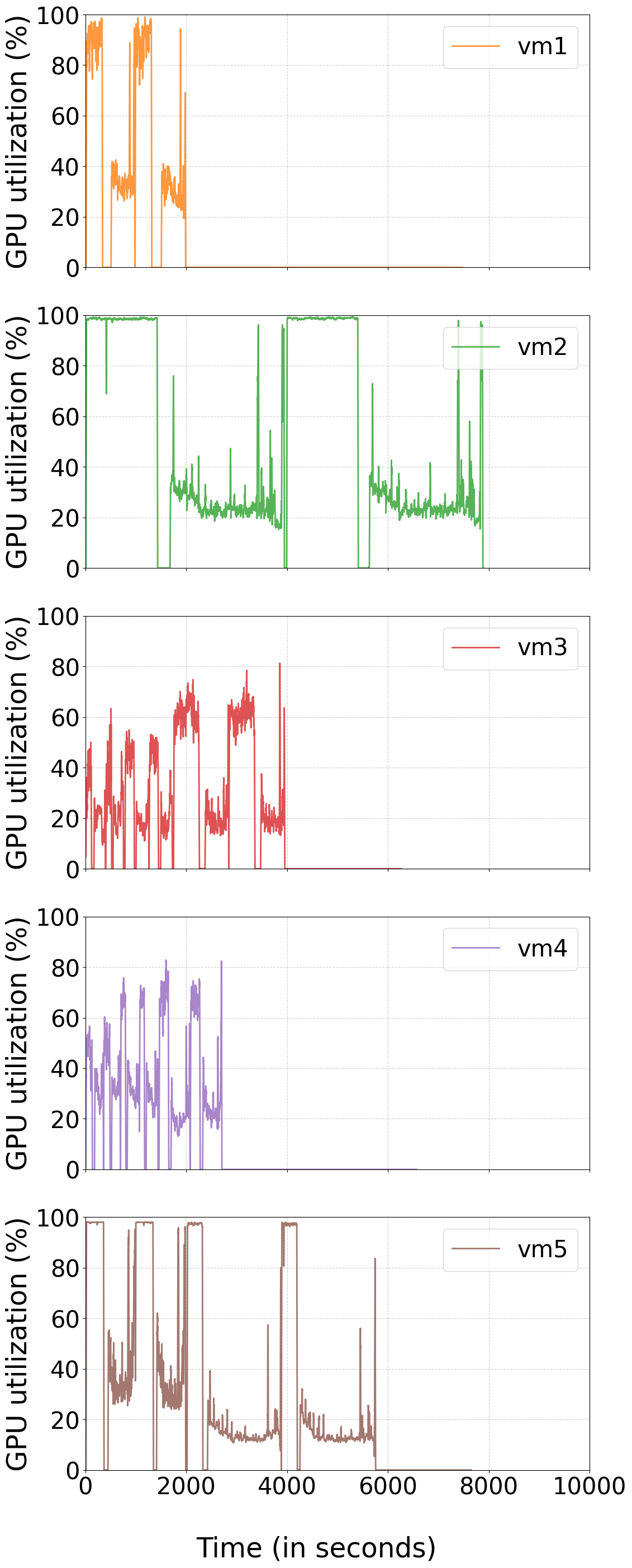} &
\hspace{-1em}
\includegraphics[width=1.70in, angle=0]{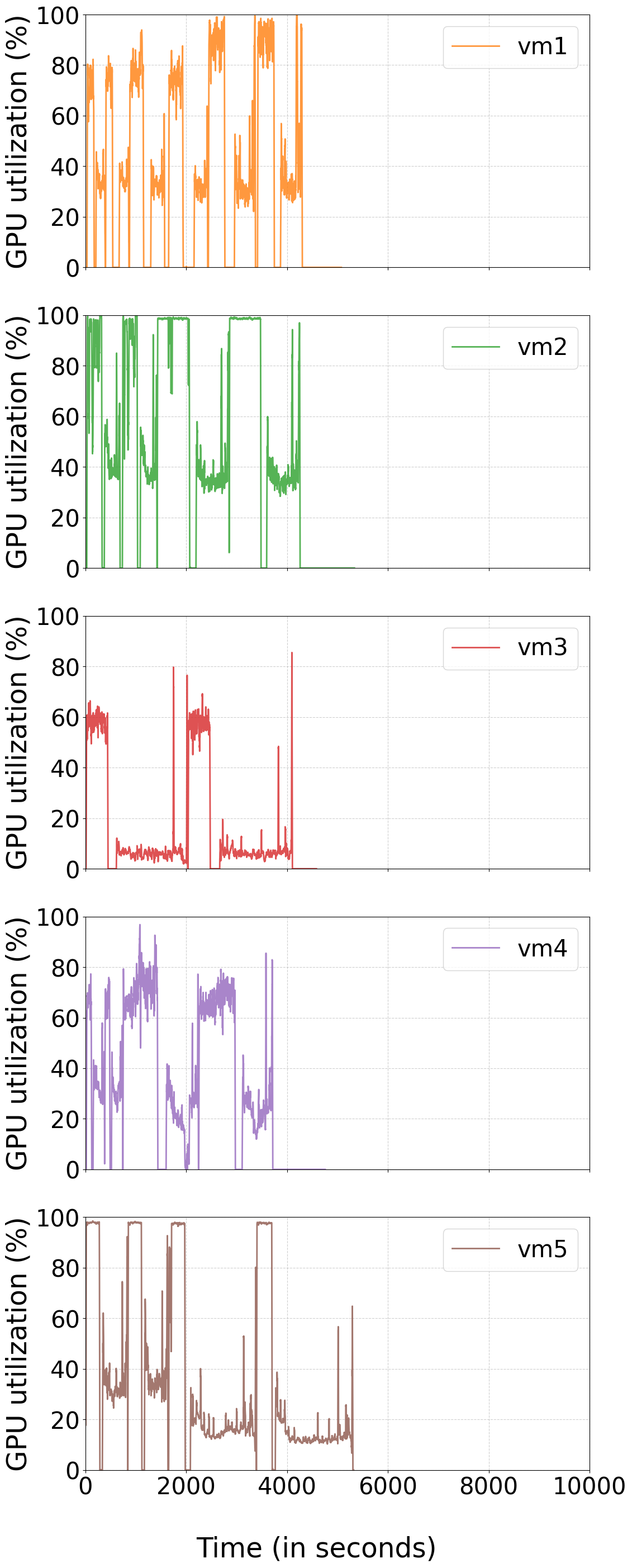} &
\hspace{-1em}
\includegraphics[width=1.70in, angle=0]{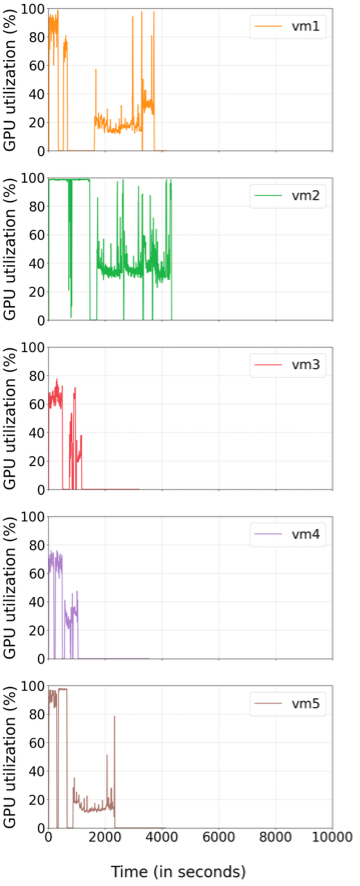} \\
(a) Greedy Strategy & (b) Dynamic Strategy & (c) FJSP Strategy (1-stage) & (d) FJSP Strategy (2-stage) \\
\end{tabular}
\caption{GPU utilization plots (for the 5 VMs) for different strategies on a representative subset}
\label{fig-gpu-utilization-plots}
\end{figure*}

We report the CPU and GPU utilization plots in this appendix for a representative subset. Figure~\ref{fig-cpu-utilization-plots} shows the CPU utilization of the different VMs for different strategies. Figure~\ref{fig-gpu-utilization-plots} shows the GPU utilization of the different VMs for different strategies. 
\end{document}